\documentstyle[12pt,epsfig,a4]{article} 
\newcommand{\vs}{\vspace{-0.25cm}}
\begin{document}
\begin{center}
{\Large{\bf Resummation of fermionic in-medium ladder diagrams to all orders}}
\bigskip

N. Kaiser\\
\medskip
{\small Physik-Department T39, Technische Universit\"{a}t M\"{u}nchen,
    D-85747 Garching, Germany

\smallskip
{\it email: nkaiser@ph.tum.de}}
\end{center}
\medskip
\begin{abstract}
A system of fermions with a short-range interaction proportional to the 
scattering length $a$ is studied at finite density. At any order $a^n$, we 
evaluate the complete contributions to the energy per particle $\bar E(k_f)$ 
arising from combined (multiple) particle-particle and hole-hole rescatterings 
in the medium. This novel result is achieved by simply decomposing the 
particle-hole propagator into the vacuum propagator plus a medium-insertion and 
correcting for certain symmetry factors in the $(n-1)$-th power of the 
in-medium loop. Known results for the low-density expansion up to and 
including order $a^4$ are accurately reproduced. The emerging series in $a k_f$ 
can be summed to all orders in the form of a double-integral over an arctangent 
function. In that representation the unitary limit $a\to \infty$ can be taken 
and one obtains the value $\xi= 0.5067$ for the  universal Bertsch parameter. 
We discuss also applications to the equation of state of neutron matter at low 
densities and mention further extensions of the resummation method.          
\end{abstract}

\bigskip

\vspace{-0.3cm}
\section{Introduction and summary}
Dilute degenerate many-fermion systems with large scattering lengths are of
interest e.g. for modeling the low-density behavior of nuclear or neutron star 
matter. Also dilute systems of ultracold atoms can nowadays be trapped. 
Because of the possibility to tune (magnetically) atomic interactions through
so-called Feshbach resonances, ultracold fermionic gases provide an 
exceptionally valuable tool to explore the (non-perturbative) many-body 
dynamics at weak and strong coupling together with the transition from the 
superconducting to the Bose-Einstein condensed state. Of particular interest 
in this context is the so-called unitary limit in which the two-body 
interaction has the critical value to support a boundstate at zero energy. 
In that special situation the s-wave scattering length diverges, $a\to \infty$, 
and the (strongly) interacting many-fermion system becomes scale invariant. Its 
groundstate energy is then determined by a single universal number, the 
so-called Bertsch parameter $\xi$, which measures the ratio of the energy per 
particle, $\bar E(k_f)^{(\infty)}$, to that of a free Fermi gas, $\bar E(k_f)^{(0)}=
3k_f^2/10M$. Here, $k_f$ denotes the Fermi momentum and $M$ stands for the 
(heavy) fermion mass.  

The calculation of $\xi$ is an intrinsically non-perturbative problem which 
has been approached in recent years by numerical quantum Monte-Carlo 
simulations (in a periodic box). As the state of the art, at present a value of 
$\xi \simeq 0.38$ seems to emerge from these calculations \cite{montecarlo}, 
presumably with still debatable error bars due to finite-size corrections etc. 
The equation of state of neutron matter at low densities has also been studied 
using quantum Monte-Carlo techniques \cite{carlson2,carlson1}. Due to the very 
large neutron-neutron scattering length, $a_{nn} \simeq 19\,$fm, neutron matter 
at low densities, $\rho_n =k_f^3/3\pi^2 \leq 0.05\,$fm$^3$, is supposed to be a
fermionic gas close to the unitarity limit. The results of a variety of
sophisticated many-body calculations are summarized in Figs.\,3,4 of 
ref.\,\cite{carlson1} and these give indications for a value of $\xi_{nn}\approx 0.5$.

On the other hand effective field theory methods have been used to rederive and
systematically improve the low-density expansion for a system of fermions with 
short-range interactions \cite{hammer,furn,hammerrev}. In particular, effective 
range corrections and (logarithmic) contributions from three-particle scattering 
have been included. In an unpublished paper, Steele \cite{steele} has computed 
several higher order contributions proportional to $(a k_f)^3$ and $(a k_f)^4$. 
These are generated either by multiple particle-particle and hole-hole rescatterings in 
the medium or they arise from particle-hole ring diagrams. Based on some 
numerical evidence, Steele argued that in the limit of large space-time 
dimensions $D$ the hole-hole contributions would be suppressed. Under this 
assumption the resummation of the particle-particle contributions becomes
possible (in the limit $D\to \infty$) in the form of a simple geometrical series: 
$-(2a k_f/3\pi)[1+2 a k_f/\pi]^{-1}$. (We are choosing the sign-convention such 
that a positive scattering length $a>0$ corresponds to attraction.) The Bertsch 
parameter following from Steele's approximation is $\xi^{(\rm St)} = 4/9$, 
surprisingly close to recent quantum Monte-Carlo results. The validity of 
Steele's arguments concerning the expansion in $1/D$ has been critically 
reassessed in the more elaborate work by Sch\"afer et al.\,\cite{schaefer}. There 
it has been shown that if the strong coupling limit $a\to \infty$ is taken after 
the limit $D\to \infty$ the universal Bertsch parameter is $\xi^{(\infty)}=1/2$ 
(see eq.(44) in ref.\,\cite{schaefer}). 

Moreover, the particle-particle ladders (for $D=4$) have been resummed in 
ref.\,\cite{schaefer} in the form of a phase space integral over a geometrical 
series, $-a[1+a k_f F_{pp}(s,\kappa)/\pi]^{-1}$, and the corresponding Bertsch 
parameter had the (rather small) value $\xi^{(pp)} \simeq 0.24$. In the same 
paper an analogous expression for the sum of all hole-hole ladders has been 
given (see eq.(25) in ref.\,\cite{schaefer}). In that case the geometrical 
series involves two subtractions (hole-hole ladders start to contribute 
at order $a^3$) and the unitary limit $a \to \infty$ does not exist for this 
term. The same feature applies to the sum of all particle-hole ring diagrams, 
which have been discussed in section IV of ref.\,\cite{schaefer}. Beyond these
partial resummation results there exist still large classes of diagrams with 
mixed particle-particle and hole-hole ladders which have not been considered in 
ref.\,\cite{schaefer} (or elsewhere in the literature). Clearly, these should 
all be resummed in order to see whether they contribute in the limit $a \to 
\infty$ to the universal Bertsch parameter $\xi$. The first mixed ($pp$ and 
$hh$) ladder appears at order $a^4$ and it has been considered by Steele 
\cite{steele}. Hammer et al.\,\cite{confirm} have carefully checked Steele's 
calculation (for $D=4$) and they find differences for just this term. In 
particular, the numerical value given for it in eq.(16) of ref.\,\cite{steele} 
(middle term) has to be corrected and multiplied by a factor 2. Actually,  
Hammer et al.\,\cite{confirm} find a simpler analytical representation for it in 
the form of a phase space integral over the squared particle-particle bubble times 
the hole-hole bubble, $3 F^2_{pp}(s,\kappa)F_{pp}(-s,\kappa)$ (see eq.(7) in 
section 3). Besides this leading term, results for mixed particle-particle and 
hole-hole ladders at higher orders $a^n, n\geq 5$ are not known. For large $n$ 
their classification and combinatorial counting is already a non-trivial 
problem.  

The purpose of the present paper is to close this gap. We will evaluate, at 
any order $a^n$, the complete contributions to the energy per particle 
$\bar E(k_f)$ arising from combined (multiple) particle-particle and hole-hole 
rescatterings in the medium. A key to the solution of this problem 
is a different organization of the many-body calculation from the start. 
Instead of treating (propagating) particles and holes separately, we keep them 
together and measure the difference to the propagation in vacuum by a 
``medium-insertion''. The latter involves a delta-function for on-shell 
kinematics and a step-function which restricts momenta to the region inside the 
Fermi sphere. In that organizational scheme the pertinent in-medium loop (or 
in-medium bubble) is complex-valued. The contribution to the energy per particle 
$\bar E(k_f)$ at order $a^n$ is therefore not obtained directly from the 
$(n-1)$-th power of the in-medium loop. However, after reinstalling the 
symmetry factors $1/(j+1)$ which belong to diagrams with $j+1$ double 
medium-insertions, a real-valued expression is retained for all $n$. Known 
results for the low-density expansion up to and including order $a^4$ are 
accurately reproduced in our scheme. The emerging series in $a k_f$ can even 
be summed to all orders in the form of a double-integral over an arctangent 
function. In that explicit representation the unitary limit $a\to \infty$ can 
be taken straightforwardly and one finds the value $\xi= 0.5067$ for the  
universal Bertsch parameter. As an application we discuss the equation of 
state of neutron matter at low densities and finally we mention further 
possible extensions of the resummation method. 
\begin{figure}
\begin{center}
\includegraphics[scale=0.4]{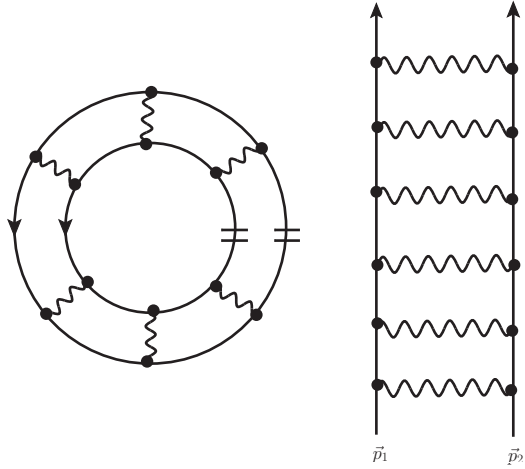}
\end{center}
\vspace{-.6cm}
\caption{Left: A closed multi-loop diagram with (at least) two 
medium-insertions representing a contribution to the energy density. Right: 
After opening at the pair of  adjacent medium-insertions (symbolized by short 
double-lines) the planar ladder diagram results. A wiggly line symbolizes the 
contact interaction proportional to the scattering length $a$.}
\end{figure} 

\vspace{-0.5cm}          
\section{Preparation: In-medium propagator}
We are interested in the equation of state of a (non-relativistic) many-fermion
system with a short-range two-body interaction proportional to the scattering 
length $a$. In perturbation theory the interaction contributions to the 
energy density are represented by closed multi-loop diagrams with a certain 
number of contact vertices. Their evaluation proceeds via Feynman rules which 
introduce a factor $4\pi a\, i/M$ for each interaction vertex and the 
(non-relativistic) particle-hole propagator:
\begin{eqnarray} G(p_0,\vec p\,) &=& i\bigg({\theta(|\vec p\,|-k_f)\over p_0-
\vec p^{\,2}/2M+i \epsilon }+{\theta(k_f-|\vec p\,|)\over p_0-\vec p^{\,2}/2M-i 
\epsilon}\bigg) \nonumber \\ &=& {i \over p_0-\vec p^{\,2}/2M+i \epsilon } 
-2\pi\, \delta(p_0-\vec p^{\,2}/2M)\,\theta(k_f-|\vec p\,|)\,, \end{eqnarray}
for an internal fermion-line carrying energy $p_0$ and momentum $\vec p$. The 
(large) fermion mass is denoted by $M$. The identity in the second line of 
eq.(1) gives the separation of the in-medium propagator $G(p_0,\vec p\,)$ into 
the vacuum propagator and a ``medium-insertion''. Note that by construction 
a medium-insertion puts a particle onto the mass-shell $p_0=\vec p^{\,2}/2M$  and 
restricts its momentum to the interior of the Fermi sphere $|\vec p\,|<k_f$. 
The Fermi momentum $k_f$ is related to the density by $\rho = g\, k_f^3/6\pi^2$. 
We are considering here only the case with spin-degeneracy factor $g=2$. 
Relevant many-body contributions to the energy density come from diagrams with 
at least two medium-insertions. In case of the closed ladder diagram shown in 
Fig.\,1 this minimal pair of medium-insertions has to placed on adjacent 
positions of the double-ring, for the following reason. After opening one gets 
a planar ladder diagram whose energy denominators are all equal to differences 
of fermion kinetic energies. Only for this planar topology the resulting factors 
of $M$ (from the energy denominators) will balance the $1/M$ factors from the 
interaction vertices such that a finite result remains (at each order $a^n$) in 
the non-relativistic limit. The open ladder diagram shown in Fig.\,1 comes in 
all possible variations with further medium-insertions on internal lines. Due 
to the momentum-independence of the contact interaction all these multi-loop 
diagrams factorize (successive loops have nothing in common) and they can therefore 
be summed together in the form of a power of the in-medium loop. 
\vspace{-0.3cm}
\section{In-medium loop generated by a contact interaction}
\begin{figure}
\begin{center}
\includegraphics[scale=0.45]{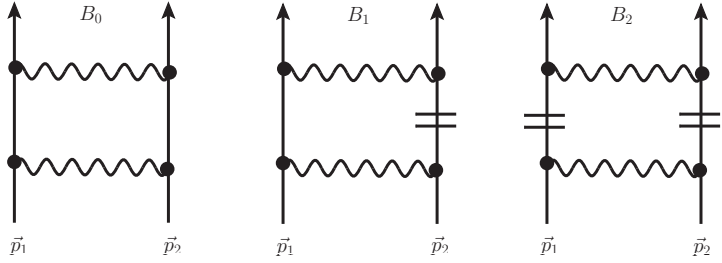}
\end{center}
\vspace{-.6cm}
\caption{The in-medium loop organized in the number of medium-insertions. The 
reflected partner of the middle diagram with one medium-insertion is not shown.
The external momenta $|\vec p_{1,2}|<k_f$ are from the region below the Fermi 
surface.}
\end{figure} 

The basic ingredient for calculating the energy density arising from ladder 
diagrams with a short-range contact interaction has been identified as the 
in-medium loop (or the in-medium bubble). As enforced by the two minimal 
medium-insertions the incoming momenta $\vec p_1$ and $\vec p_2$ are from the 
region below the Fermi surface $|\vec p_{1,2}|<k_f$. It is convenient to 
introduce their half sum $\vec P = (\vec p_1+\vec p_2)/2$ and half difference 
$\vec q = (\vec p_1-\vec p_2)/2$. In our ordering scheme the in-medium loop is 
built up from contributions with zero, one, and two medium-insertions: 
$B_0+B_1+B_2$. The corresponding one-loop diagrams are shown in Fig.\,2 and 
their evaluation is exhibited now in detail.
    
The contribution $B_0$ with zero medium-insertions is the well-known 
rescattering bubble in vacuum. It is normalized relative to the tree-level 
interaction and therefore it includes only one factor $4\pi a\, i/ M$. The 
evaluation of the loop integral proceeds as follows: 
\begin{eqnarray}B_0 &=& { 4\pi a\, i \over M} \int\limits_{-\infty}^\infty \!
{dl_0 \over 2\pi} \int \!{d^3 l \over (2\pi)^3} {i \over ( \vec p_1^{\,2}+\vec 
p_2^{\,2})/4M+l_0-(\vec P+\vec l\,)^2/2M+i \epsilon }\nonumber \\ && \times  
{i \over ( \vec p_1^{\,2}+ \vec p_2^{\,2})/4M-l_0-(\vec P-\vec l\,)^2/2M+i 
\epsilon } \nonumber \\ &=&  4\pi a \int\! {d^3 l \over (2\pi)^3} \,{1 \over  
\vec l^{\,2}-\vec q^{\,2}-i \epsilon }= {2a \over \pi } \int\limits_0^\infty\! dl 
\,\bigg( 1 + {\vec q^{\,2} \over  l^2 -\vec q^{\,2}-i \epsilon } \bigg)  = 0 + 
i\, a |\vec q \,|\,,   \end{eqnarray}
using residue calculus for the energy integral $\int_{-\infty}^\infty \!dl_0$ and 
in the last step the rule $\int_0^\infty dl\, 1 = 0$ of dimensional 
regularization has been applied. In other regularization schemes the occurring 
scale-dependent constant $-2\mu/\pi$ is absorbed into $a^{-1}$ to define the 
renormalized (physical) scattering length \cite{steele,schaefer}. The 
resummation of infinitely many rescatterings in the vacuum in the form of a 
geometrical series leads to the unitarized scattering length approximation:
\begin{equation}  f= a\Big\{ 1+ i a |\vec q \,|+ (i a |\vec q \,|)^2 + \dots \Big\} 
= {1\over a^{-1}-i |\vec q \,|} = {1\over |\vec q \,|(\cot \delta_0 -i)} \,,
\end{equation} 
which implies the relation  $\tan \delta_0 = a |\vec q \,|$ for the s-wave 
scattering phase shift $\delta_0$.
\begin{figure}
\begin{center}
\includegraphics[scale=0.55]{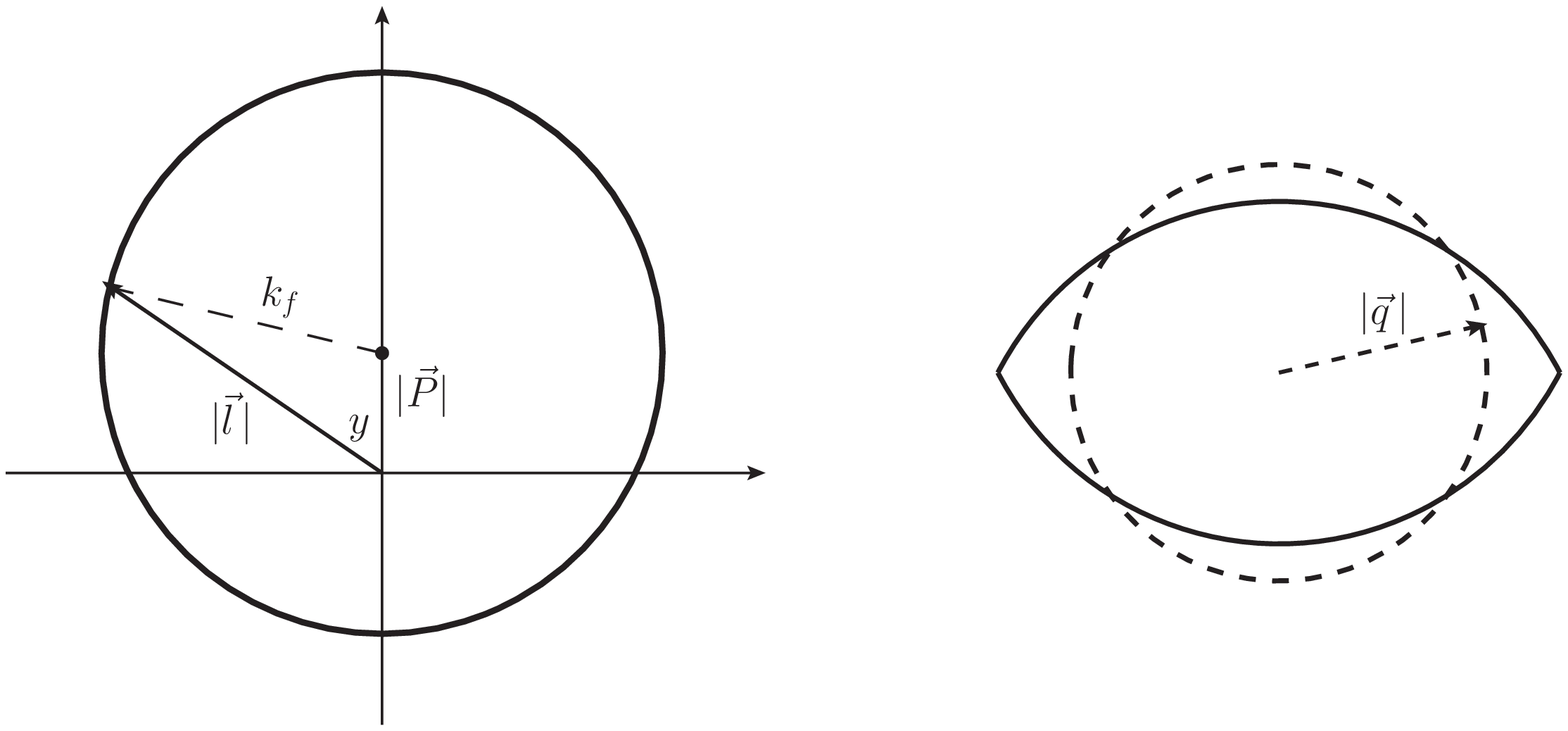}
\end{center}
\vspace{-.5cm}
\caption{Integration regions in momentum space for the real and imaginary part 
of the in-medium loop.}
\end{figure}

Next, we turn to the term $B_1$ from diagrams with one medium-insertion. 
There are two equal contributions which differ in their representation merely 
by the sign of the loop momentum $\vec l $ and together they read:
\begin{equation} B_1 =   -4\pi a \int {d^3 l \over (2\pi)^3} {1 \over  
\vec l^{\,2}-\vec q^{\,2}-i \epsilon } \Big\{ \theta(k_f-|\vec P-\vec l\,|)+ 
\theta(k_f-|\vec P+\vec l\,|)\Big\} \,.\end{equation} 
The integral over a shifted Fermi sphere is most suitably performed by using 
spherical coordinates  with the range $0< |\vec l\,| < |\vec P|y +\sqrt{
k_f^2 -\vec P^{\,2}(1-y^2)}$ of the magnitude $|\vec l\,|$. The quantity $y$ is a 
directional cosine which covers the full range $-1<y<1$ since the displacement 
vector $\vec P$ satisfies the condition $|\vec P|<k_f$ (see also the left part 
of Fig.\,3). Carrying out this procedure one obtains the following result for the 
real part:
\begin{equation} {\rm Re}\,B_1 = -{a k_f \over \pi} \, R(s,\kappa) \,,
\end{equation}   
with the logarithmic function: 
\begin{equation} R(s,\kappa) = 2 +{1\over 2s}[1+(s+\kappa)^2]\ln{1+s +\kappa
\over |1-s -\kappa|}+{1\over 2s}[1+(s-\kappa)^2]\ln{1+s -\kappa
\over 1-s +\kappa}\,, \end{equation}
written in terms of the two dimensionless variables $s = |\vec p_1+ \vec p_2|
/2k_f$ and $\kappa= |\vec p_1 -\vec p_2|/2k_f$. Since both external momenta 
$\vec p_{1,2}$ are from inside the Fermi sphere one has the additional 
constraint $s^2+\kappa^2<1$. It is worth to note that the function $R(s,\kappa)$ 
in eq.(6) is equal to the sum of the particle-particle bubble and the 
hole-hole bubble, $R(s,\kappa)=F_{pp}(s,\kappa)+F_{pp}(-s,\kappa)$, with the 
unusual feature that the latter is also taken at momenta below the Fermi 
surface. For comparison the particle-particle bubble reads 
\cite{steele,schaefer}:
\begin{equation} F_{pp}(s,\kappa) = 1+s -\kappa \ln{1+s +\kappa \over |1+s 
-\kappa|}+{1\over 2s}(1-s^2-\kappa^2)\ln{|(1+s)^2 -\kappa^2 |\over 1-s^2 -
\kappa^2}\,. \end{equation}
Finally, we come to the right diagram in Fig.\,2 with two-medium insertions. 
Obviously, it generates a purely imaginary contribution. The total imaginary 
part of the in-medium loop has the following representation:   
\begin{eqnarray} {\rm Im}(B_0+B_1+B_2) &=& 4 \pi a  \int {d^3 l \over (2\pi)^3} \,
\pi\, \delta(\vec l^{\,2}-\vec q^{\,2}) \nonumber \\ && \times \bigg\{\Big[1-
\theta(k_f-|\vec P-\vec l\,|)\Big]\Big[ 1-\theta(k_f-|\vec P+\vec l\,|)\Big]
\nonumber \\ && \quad +\theta(k_f-|\vec P-\vec l\,|)\, \theta(k_f-|\vec P+
\vec l\,|) \bigg\} \,,\end{eqnarray}
where we have suitably arranged terms with no, one, and two step-functions 
$\theta(...)$. The first term in eq.(8) of the form $[1-\theta(...)][1-
\theta(...)]$ makes no contribution to the imaginary part since the 
corresponding phase space is completely Pauli-blocked: $2k_f^2<(\vec P-\vec l\,
)^2+(\vec P+\vec l\,)^2 = 2(\vec l^{\,2}-\vec q^{\,2})+\vec p_1^{\,2}+\vec p_2^{\,2}
<2k_f^2$. This equation expresses the obvious fact that on-shell scattering of 
two particles from below the Fermi surface into the region above the Fermi 
surface is not possible due to energy conservation. Thus there remains the 
imaginary part due to the second $\theta(...)\,\theta(...)$ term in eq.(8). After 
visualizing the occurring product of $\theta$- and $\delta$-functions one sees 
that the imaginary part has a nice geometrical interpretation: namely as 
$|\vec q\,|$ times that part of the solid angle of a (centered) sphere of radius 
$|\vec q\,|$ which lies inside the intersection region of two spheres of radius 
$k_f$ with their centers displaced by $2|\vec P|$. The corresponding 
configuration of spheres is sketched in the right part of Fig.\,3. After 
inclusion of the appropriate prefactor the result for the imaginary part of the 
in-medium loop reads:
\begin{equation} {\rm Im}(B_0+B_1+B_2) = {B_2 \over 2i} = a k_f \, 
I(s,\kappa)\,,  \end{equation}
with the (non-smooth) function:
\begin{equation}I(s,\kappa)=  \left\{ \begin{array} {cll} \kappa  &
\rm {for} & 0<\kappa<1-s \,, \\ 
\displaystyle {1 \over 2s}(1-s^2-\kappa^2)& \rm {for} & 1-s<
\kappa<\sqrt{1-s^2}\,, \\ \end{array} \right. \end{equation}
where $\kappa$ lies in the interval $0< \kappa<\sqrt{1-s^2}$. It is interesting 
to observe that the diagram with two medium-insertions alone gives twice the 
total imaginary part. Putting the real and imaginary pieces together the 
complex-valued in-medium loop reads:
\begin{equation} B_0+B_1+B_2=  -{a k_f \over \pi} \,\Big\{ R(s,\kappa)- i \pi \,
I(s,\kappa)\Big\}  \,, \end{equation}  
and if the contribution from the diagram with two medium-insertions is taken out, 
the imaginary part of that same expression changes sign:  
\begin{equation} B_0+B_1=  -{a k_f \over \pi} \,\Big\{ R(s,\kappa)+ i \pi \,
I(s,\kappa)\Big\}  \,.\end{equation}
Actually, this special property of the in-medium loop turns out to be crucial in 
order to derive the correct expression for the energy per particle $\bar E(k_f)$ 
from powers of the (complex-valued) in-medium loop. The derivation of the energy 
per particle is the topic of the next section.
\vspace{-0.3cm}   
\section{Energy per particle}
\begin{figure}
\begin{center}
\includegraphics[scale=0.4]{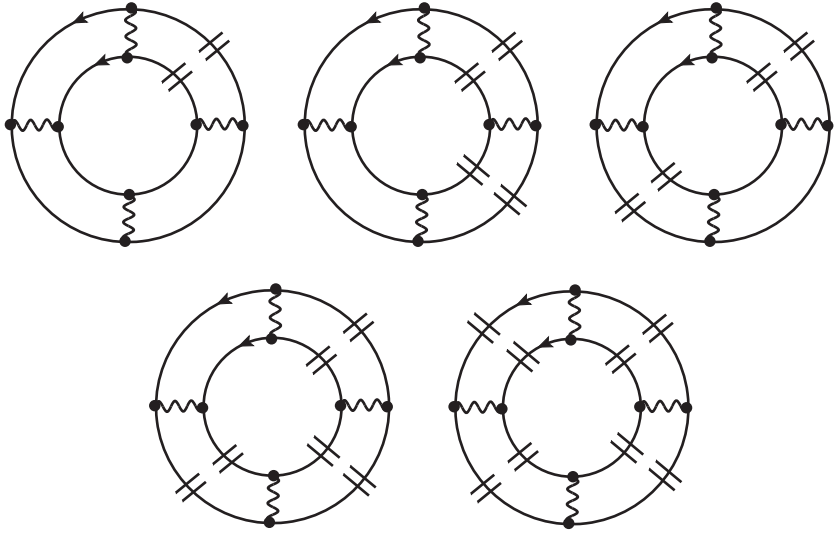}
\end{center}
\vspace{-.5cm}
\caption{In-medium diagrams contributing to the energy density at order 
$a^4$. Taking into account the proper symmetry factors, their total sum is 
given by the real-valued expression: $(R+i \pi I)^3+(R+i \pi I)^2(-2\pi i I)
(1+1/2)+ (R+i \pi I)(-2\pi i I)^2+(-2\pi i I)^3/4 =R(R^2-\pi^2 I^2)$. Diagrams 
with a single medium-insertion on the inner or outer fermion-line between 
consecutive interactions are not shown.}
\end{figure}
In this section we show how the contributions to the energy per particle $\bar 
E(k_f)$ at any order $a^n$ can be constructed from the in-medium loop 
and how the emerging series in $a k_f$ can be summed to all orders. Consider the 
(open) ladder diagram with $n$ contact interactions. It is given by the 
$(n-1)$-th power of the in-medium loop times a factor $4\pi a\,i/M$. Closing the 
two open fermion-lines introduces an integration over the allowed phase space  
$|\vec p_{1,2}|< k_f$. The emerging integrand $(R-i \pi I)^{n-1}$ for the energy 
density at order $a^n$ would be complex-valued and therefore it cannot 
yet be the correct one. The deficit of the (naive) iteration method at this 
intermediate stage becomes evident if one draws all diagrams with (repeated) 
pairs of adjacent medium-insertions. The corresponding set of diagrams at fourth 
order $a^4$ is shown in Fig.\,4. These (decorated) diagrams have additional 
symmetry factors which are not respected by the binomial expansion of $(R-i \pi 
I)^{n-1} = [(R+i \pi I)+(-2i \pi I)]^{n-1}$. Furthermore, the same symmetry factors 
correct for the overcounting of certain diagrams as introduced by the terms of 
the binomial series. As a result of this combinatorial analysis one has to 
reweight in the binomial series expansion of $[(R+i \pi I)+(-2i \pi I)]^{n-1}$ 
the $j$-th power of $-2i \pi I$ coming from the diagrams with repeated double 
medium-insertions with the appropriate symmetry factor $1/(j+1)$. This crucial 
amendment leads to the following summation formula:  
\begin{equation} \sum_{j=0}^{n-1}  (R+i \pi I)^{n-1-j}(-2i \pi I)^j {n-1 \choose j}
{1\over j+1} = {1\over 2i \pi I n}\Big\{ (R+i \pi I)^n- (R-i \pi I)^n\Big\} 
\,. \end{equation}   
Note that the identity ${n-1 \choose j}/(j+1)= {n \choose j+1}/n$  introduces the 
binomial coefficients for the $n$-th power of a sum,  and in this way one can easily 
reproduce the result on the right hand side of eq.(13). In the diagrammatic 
representation of the energy density, ${n \choose j+1}$ is the number of different
possibilities to attach $j+1$ double medium-insertions on a ring with 
$n$ segments and the additional factor $1/n$ comes from the $n$ rotations which  
transform the ring into itself. 

Inspection of the right hand side of eq.(13) shows that the resulting 
homogeneous polynomials of degree $n-1$ in $R$ and $\pi I$ are manifestly real 
for all $n$. The first five terms read:  $n=1$: $1$, $n=2$: $R$,  $n=3$: 
$R^2-\pi^2I^2/3$, $n=4$: $R(R^2-\pi^2I^2)$, $n=5$: $R^4-2R^2\pi^2I^2+\pi^4I^4/5$. \\
In Fig.\,4 and the appended caption it is shown explicitly how the real-valued 
expression $R(R^2-\pi^2I^2)$ results from the sum of (complex-valued) diagrams 
once their symmetry factors are taken into account. The same graphology gives at 
second order $a^2$: $ (R+i \pi I)+ (-2i \pi I)/2=R$, and at third order $a^3$: 
$ (R+i \pi I)^2+(R+i \pi I)(-2i \pi I)+(-2i \pi I)^2/3=R^2-\pi^2 I^2/3$.  

At this point we have achieved a representation which includes, at any order 
$a^n$, the complete contributions from all mixed particle-particle and hole-hole 
ladders. One can even go further and sum up the whole series of ladder diagrams 
to all orders. The pertinent series $\sum_{n=1}^\infty [-a k_f(R\pm i \pi I)/\pi
]^n/n$ can be solved easily by a (complex) logarithm. After inclusion of the 
kinetic energy and the Fock exchange-term (which reduces the Hartree term 
discussed so far by a factor $1-1/g=1/2$) the complete expression for the energy 
per particle reads:
\begin{equation} \bar E(k_f)= {k_f^2 \over 2M} \Bigg\{ {3\over 5}-{48\over \pi} 
\int\limits_0^1 \!ds\, s^2  \!\!\int\limits_0^{\sqrt{1-s^2}}  \!\!d\kappa \, 
\kappa \, \arctan { a k_f\, I(s,\kappa) \over 1+ \pi^{-1}ak_f \,R(s,\kappa)}\Bigg\} 
\,. \end{equation}
The occurring arctangent function refers to the usual branch with odd parity,  
$\arctan(-x) = - \arctan x$, and values in the interval $[-\pi/2, \pi/2]$. 
Other branches of the arctangent function are excluded by the weak coupling 
limit $a\to 0$, which has to give zero independent of the sign of the scattering 
length $a$. Note that there occurs a discontinuity (by an amount $-\pi$) when 
the denominator passes through zero from positive to negative values. This 
happens for all positive values of $a k_f$ since the function $R(s,\kappa)$ has 
a logarithmic singularity at $s=0,\,\kappa=1$. The integral-representation for 
$\bar E(k_f)$ given in eq.(14) has some similarity with the expression one 
obtains from the resummed particle-hole ring diagrams (see eq.(29) in 
ref.\,\cite{schaefer}). In that case the arctangent function involves two 
subtractions (particle-hole ring diagrams start to contribute at order $a^3$) 
and the integral extends over a 4-momentum transfer $\int_0^\infty dq_0 
\int_0^\infty dq\,q^2$. In the present case the reduction to a double-integral 
(over the quarter unit-disc) has been obtained by employing the following master 
formula for integrals over the interior of two Fermi spheres: 
\begin{equation} \int\limits_{|\vec p_{1,2}|<k_f}\!\!\!{d^3 p_1d^3p_2\over(2\pi)^6} \,
F(s, \kappa) = {2k_f^6 \over \pi^4} \int\limits_0^1 \! ds\, s^2 \!\!\int
\limits_0^{\sqrt{1-s^2}}  \!\!d\kappa \, \kappa \, I(s,\kappa) F(s, \kappa)\,, 
\end{equation}      
where $s = |\vec p_1+ \vec p_2|/2k_f$ and $\kappa= |\vec p_1 -\vec p_2|/2k_f$.
Surprisingly, the imaginary part function $I(s,\kappa)$ defined in eq.(10) 
occurs here as the pertinent weighting function. Integrals over the outer region 
of two Fermi spheres can be reduced in a similar way: 
\begin{eqnarray} \int\limits_{|\vec p_{1,2}|>k_f}\!\!\!{d^3 p_1 d^3p_2 \over 
(2\pi)^6}\, F(s, \kappa) \, \theta(1-s)\!\!\!&=&\!\!\!{k_f^6 \over \pi^4} \int
\limits_0^1 \! ds\, s  \!\!\int\limits_{\sqrt{1-s^2}}^\infty  \!\!d\kappa \,\kappa 
\, F(s, \kappa)\nonumber \\ & & \times \Big[(s^2+\kappa^2-1)\, \theta(1+s-
\kappa) +2s \kappa \, \theta(\kappa-1-s)\Big]\,. \end{eqnarray} 
This formula is useful e.g. for evaluating the contributions to the energy per 
particle which arise from multiple hole-hole rescatterings in the medium 
\cite{steele,schaefer}. 
\vspace{-0.3cm}
\section{Expansion in powers of $\boldmath{a k_f}$}
Several orders in the low-density expansion of the energy per particle $\bar 
E(k_f)$ are known \cite{hammer}. The terms stemming from particle-particle 
and hole-hole ladders can be used as a check of our calculation which is 
organized differently by not treating separately particles and holes. We find up 
to and including fourth order:     
\begin{eqnarray} \bar E(k_f)&=& {k_f^2 \over 2M} \bigg\{ {3\over 5}-{2\over 
3\pi} a k_f+{4\over 35 \pi^2}(11-2\ln 2)a^2 k_f^2 \nonumber \\ & & -0.0755733  
\, a^3 k_f^3 + 0.0524813 \,a^4 k_f^4 +  \dots \bigg\}\,. \end{eqnarray} 
As it must be, the linear and quadratic coefficients agree analytically. 
Since only double-integrals are involved the other numerical coefficients 
can be obtained with high precision. One finds again good agreement 
with existing calculations \cite{hammer,steele,confirm}. The third order 
coefficient is: $0.0861836-0.0106103 = 0.0640627+0.0115106$, where the numbers on 
the left hand side correspond to our separation $R^2-\pi^2 I^2/3$ and those 
on the right side refer to the sum of two-fold particle-particle and two-fold 
hole-hole rescatterings. The same comparison for the fourth order coefficient 
gives: $0.0671902-0.0147089 = 0.0383116 -0.0006851+6 \cdot 0.0024758$, with our 
separation $R^3- R\pi^2I^2$ versus the sum of triple  particle-particle, triple 
hole-hole, and combined particle-particle and hole-hole scatterings.  At this 
point it is important to note that a factor of $2$ is missing in Steele's numerical 
result for the latter contribution (see middle term in eq.(16) of 
ref.\,\cite{steele}). This error has been confirmed by Hammer et al.\,\cite{confirm}
who have carefully checked Steele's calculation. It has become evident that the 
squared imaginary part $I(s,\kappa)^2$ of the in-medium loop includes also 
important many-body correlation effects. Their precise mapping into the 
(traditional) particle-hole counting scheme is not obvious.

As mentioned in the introduction, Steele \cite{steele} has suggested a resummation 
to all orders in form of a simple geometrical series:
\begin{equation} \bar E(k_f)^{(\rm St)}=   {k_f^2 \over 2M} \bigg\{ {3\over 5} 
-{2a k_f \over 3\pi +6 a k_f} \bigg\}\,. \end{equation} 
Although the original arguments (via a $1/D$-expansion) have been (partly) 
invalidated \cite{schaefer}, it may serve as a useful reference, in particular 
since the associated Bertsch parameter $\xi^{(\rm St)}=4/9$ comes out fairly 
realistic. We adapt the expansion of the energy per particle in powers of $a k_f$:
\begin{equation} \bar E(k_f) =   {k_f^2 \over 2M} \bigg\{ {3\over 5} 
+ \sum_{n=1}^\infty {c_n \over 3}\Big(\!-{2 \over \pi}a k_f\Big)^n \bigg\}\,, 
\end{equation} 
such that all coefficients $c_n$ become $1$ for Steele's resummation result. 
Performing the same expansion with our complete expression for $\bar E(k_f)$ given
in eq.(14) we find the following values for the first dozen expansion 
coefficients $c_n$: 
\begin{eqnarray} && c_1 = 1\,, \qquad c_2 = {3\over 35}(11-2 \ln 2) = 
0.8240319119 \,, \nonumber \\  && c_3 ={\pi^2 \over 12}+{9\over 160} 
+\delta c_3= 0.8787170548\,, \qquad \delta c_3= 2.1365 \cdot 10^{-8}\,,  
\nonumber \\  && c_4 =1.22717534-{\pi^2 \over 70}\bigg(10-\pi^2
-4\ln^2 2 +{16\over 3} \ln2\bigg) = 0.958529\,, \nonumber \\ 
&& c_5 = 1.14589\,, \qquad c_6 =  1.37081\,, \nonumber \\ 
&& c_7 = 1.76240\,, \qquad c_8 =  2.19993\,, \nonumber \\ 
&& c_9 = 3.03120\,, \qquad c_{10} =  3.74458\,, \nonumber \\ 
&& c_{11} = 5.85642\,, \qquad c_{12} = 5.96732\,.  
 \end{eqnarray} 
One observes that the first few coefficients stay below $1$ while the higher 
ones show a tendency to increase appreciably with $n$. The apparent region of 
``convergence'' of the power series in $a k_f$ can therefore roughly be 
estimated as $k_f \leq 1.3/|a|$. Despite various attempts we have not succeeded  
to derive the exact value of $c_3$ in terms of (reasonable) mathematical 
constants. The value $c_3\simeq \pi^2/12+9/160 $ represents an extremely accurate 
analytical approximation, which is far more precise than all previous 
determinations of this coefficient \cite{hammer,steele}.     

For large scattering lengths $a$ the low-density expansion is of very limited 
validity. In that case one has to take the full expression for the energy per 
particle $\bar E(k_f)$ as given by eq.(14). The solid line in Fig.\,5 shows its 
ratio to the (free) Fermi gas energy $3k_f^2/10M$ as a function of the 
dimensionless parameter $a k_f$. The behavior of this ratio is demonstrated for 
both signs of the scattering length ($a>0$ for attraction and $a<0$ for 
repulsion). Outside the region $k_f |a|>6$ the curve is almost flat while a sharp 
peak develops inside with a maximum value of about $1.62$ at $ak_f \simeq -0.9$. 
The dashed curve in Fig.\,5 corresponds to Steele's \cite{steele} resummation via a 
geometrical series eq.(18). The behavior on the attractive side ($a>0$) is quite
similar while on the repulsive side  ($a<0$) the artificial pole at 
$ak_f = -\pi/2$ causes  essential differences.   
\begin{figure}
\begin{center}
\includegraphics[scale=0.46]{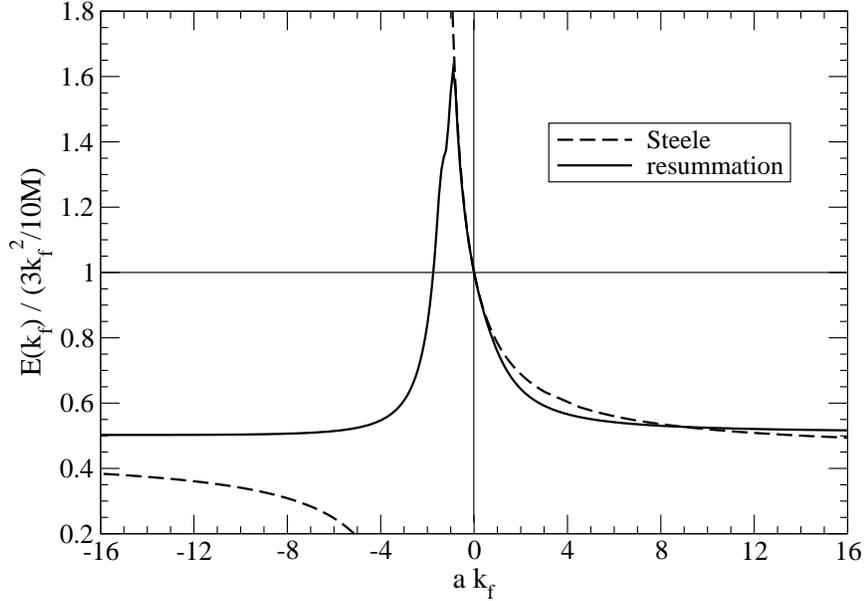}
\end{center}
\vspace{-.4cm}
\caption{Energy per particle $\bar E(k_f)$ divided by the Fermi gas energy 
$3k_f^2/10M$ as a function of the dimensionless parameter $a k_f$. An 
attractive (repulsive) contact interaction corresponds to a positive (negative)
value of the scattering length $a$.}
\end{figure}
\vspace{-0.3cm}
\section{Unitary limit}
The unitary limit $a\to \infty$ is of special interest since in this limit the 
strongly interacting many-fermion system becomes scale invariant. The energy
per particle is then determined by just a constant multiple of the (free) 
Fermi gas energy:
\begin{equation} \bar E(k_f)^{(\infty)}= {3k_f^2 \over 10 M}\, \xi \,, 
\end{equation}   
with $\xi$ the so-called Bertsch parameter. Returning to the expression for the 
energy per particle $\bar E(k_f)$ given in eq.(14) one sees that the unitary 
limit  $a\to \infty$ can be performed straightforwardly. The formula for 
calculating the  Bertsch parameter $\xi$ reads:
\begin{equation} \xi = 1- {80\over \pi} \int\limits_0^1 \!ds\, s^2\!\!\int
\limits_0^{\sqrt{1-s^2}}  \!\!d\kappa \, \kappa \, \arctan { \pi \,I(s,\kappa) 
\over R(s,\kappa)} = 0.5067\,. \end{equation}
The resulting numerical value $\xi=0.5067$ is to be compared with $\xi^{(pp)} 
\simeq 0.237$ obtained by Sch\"afer et al.\,\cite{schaefer} from the resummation 
of particle-particle ladders. Actually, $ \xi^{(pp)}$ is defined by the principal 
value integral  $ \xi^{(pp)}=1-80 \int_0^1\! ds\, s^2 -\!\!\!\!\!
\int_0^{\sqrt{1-s^2}} \! d\kappa \,\kappa \, I(s,\kappa)F_{pp}^{-1}(s,\kappa)$ and the 
necessity to treat the pole-singularity limits the precision in the numerical 
computation of this number. One observes that the additional mixed 
particle-particle and hole-hole ladders increase the Bertsch parameter $\xi$ by 
more than a factor $2$. Moreover, the divergence of the subset of hole-hole 
ladders (for $a\to \infty$) encountered in ref.\,\cite{schaefer} has disappeared. 
Clearly, the value $\xi=0.5067$ as obtained here via an analytical calculation rooted 
in perturbation theory is still considerably larger than the result $\xi \simeq 0.38$ 
from recent quantum Monte-Carlo simulations \cite{montecarlo}. 

In the work by Haussmann et al.\,\cite{zwerger} a self-consistent, 
thermodynamically consistent treatment of the unitary Fermi gas at finite 
temperatures has been presented. In their non-perturbative approach the exact one- and 
two-particle Green functions serve as an infinite set of variational parameters and 
extensive numerical work enters into the solutions of the stationarity constraints and 
the thermodynamical potentials \cite{zwerger}. The resulting Bertsch parameter at zero 
temperature $T=0$ was found to be $\xi\simeq 0.36$. This value is remarkably close to 
the experimental determinations by Bartenstein et al.\,\cite{bartenstein}, $\xi = 0.32 
\pm 0.11$, and Bourdel et al.\,\cite{bourdel}, $\xi = 0.36 \pm 0.15$. 
\vspace{-0.3cm}
\section{Application to neutron matter and outlook}
As an application of our analytical result eq.(14) for the complete resummation of 
in-medium ladder diagrams we consider the equation of state of neutron matter. Due 
to the very large neutron-neutron scattering length $a_{nn} = (18.95 \pm  0.40)
\,$fm \cite{gonzales,chen} neutron matter at low densities is supposed to be a 
Fermi gas close to the unitary limit.  Recent quantum Monte Carlo simulations 
\cite{carlson2,carlson1} based e.g. on the Argonne nucleon-nucleon potential 
give some indication for such a behavior. 
\begin{figure}
\begin{center}
\includegraphics[scale=0.4]{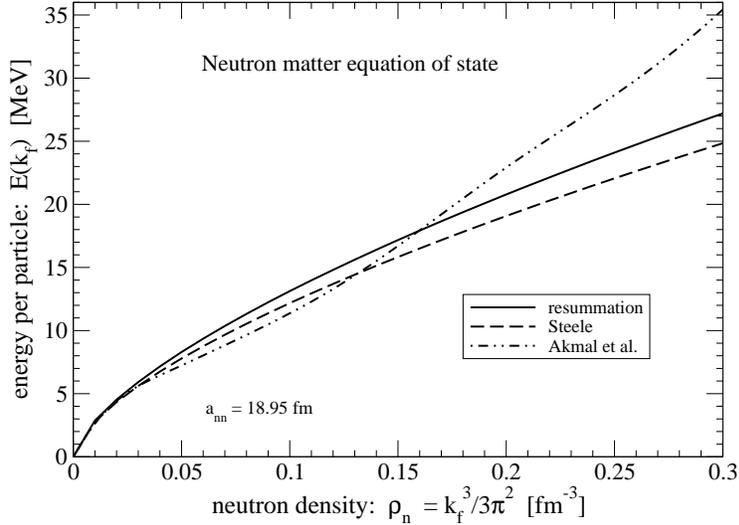}
\end{center}
\vspace{-.5cm}
\caption{Energy per particle of neutron matter versus the neutron density 
$\rho_n=k_f^3/3\pi^2$. The dash-dotted line stems from the sophisticated 
many-body calculation of ref.\cite{akmal}.}
\end{figure}

In Fig.\,6 we show the energy per particle $\bar E(k_f)$ of neutron matter as a 
function of the neutron density $\rho_n=k_f^3/3\pi^2$. The solid line results from 
our analytical formula eq.(14) inserting $a=a_{nn}=18.95$\,fm for the scattering 
length and $M=M_n = 939.57\,$MeV for the fermion mass. The dash-dotted line 
corresponds to the sophisticated many-body calculation by the Urbana group 
\cite{akmal}, to be considered as representative of realistic neutron matter 
calculations. The dashed line in Fig.\,6 reproduces Steele's suggestion eq.(18) in 
the form of a simple geometrical series. One observes good agreement up to rather high 
neutron densities of $\rho_n \simeq 0.2\,$fm$^{-3}$, where the dimensionless 
parameter $a_{nn} k_f$ reaches values up  to $a_{nn} k_f \simeq 34.3$. At higher 
neutron densities repulsive effects from three-nucleon forces (which are included 
in the Urbana calculation \cite{akmal}) start to play a more significant role. The 
inclusion of the effective range $r_{nn} = (2.75 \pm  0.11)\,$fm for s-wave 
$nn$-scattering can also lead to sizeable changes in the equation of state, as 
demonstrated in ref.\cite{schwenk}. Note however, that only the resummed 
particle-particle ladders have been considered in that work.    

Fig.\,7 shows the neutron matter equation of state in a different representation by 
plotting the ratio of the energy per particle $\bar E(k_f)$ to the (free) Fermi gas 
energy $3k_f^2/10M$ against the dimensionless parameter $a_{nn} k_f$. The dots 
in this figure reproduce results of various (quantum Monte-Carlo) calculations of 
low-density neutron matter and have been taken over from Figs.\,3,4 in 
ref.\cite{carlson1} (see original references therein). The 
solid and dashed curves are the same as in the right part of Fig.\,5 only continued 
further out in the parameter  $a k_f$.

\begin{figure}
\begin{center}
\includegraphics[scale=0.41]{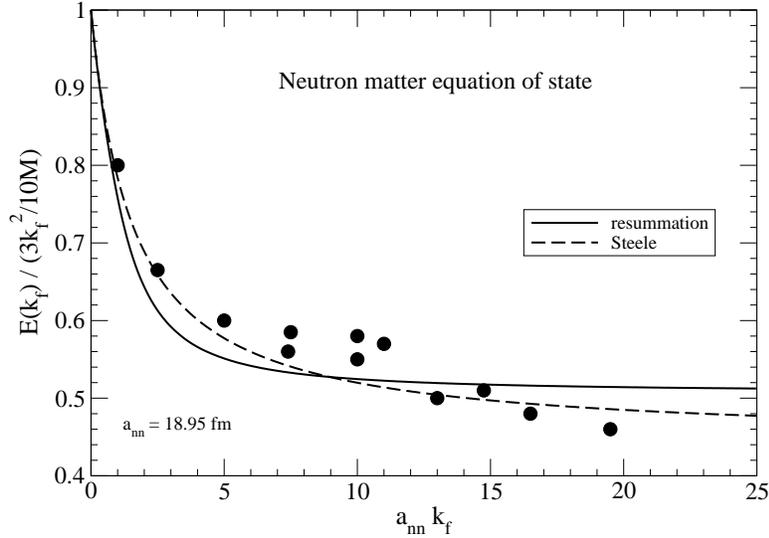}
\end{center}
\vspace{-.3cm}
\caption{Energy per particle of neutron matter divided by the Fermi gas energy 
$3k_f^2/10M$. The dots representing various (quantum Monte-Carlo) calculations 
are taken from ref.\cite{carlson1}.}
\end{figure}
The complete resummation of ladder diagrams (with a short-range contact interaction 
proportional to the s-wave scattering length $a$) as achieved in this work suggests 
numerous possible extensions. The s-wave effective range parameter and the p-wave 
scattering volumes should be included via ${\cal O}(p^2)$ terms in the contact 
interaction. Asymmetries with respect to the spin and/or isospin degrees of 
freedom can be studied via an appropriate modification of the medium-insertion. 
The generalization of the resummation method to finite temperatures would be 
an equally interesting project. Work along these lines in progress.      
\vspace{-0.3cm}
\section*{Acknowledgements}
I thank J.W. Holt, A. Schwenk and W. Weise for many  useful discussions. I thank 
H.W. Hammer for communicating to me results from own unpublished work which have been 
very valuable. This work is partially supported by the DFG Excellence Cluster ``Origin 
and Structure of the Universe''.

\end{document}